\documentclass[]{spie}  
\usepackage[]{graphicx}
\begin{document}

\title{Design of a CZT Gamma-Camera for GRB and Fast Transient
Follow-up: a Wide-Field-Monitor for the EDGE Mission
} 


\author{L. Natalucci\supit{a}, M. Feroci\supit{a}, E.Quadrini\supit{b}, P. Ubertini\supit{a}, L. Piro\supit{a}, J.W. den Herder\supit{c},
D. Barret\supit{d}, L. Amati\supit{e}, C.Budtz-Jorgensen\supit{f}, 
E. Caroli\supit{e}, S. Di Cosimo\supit{a}, M. Frutti\supit{a},  
C. Labanti\supit{e}, F.~Monzani\supit{g}, J.M.~Poulsen\supit{g}, 
L. Nicolini\supit{g}, A.~Stevoli\supit{g}
\skiplinehalf
\supit{a}INAF-Istituto di Astrofisica Spaziale e Fisica Cosmica, 
Sezione di Roma, via Fosso del Cavaliere 100, 00133 Roma, 
Italy; \\
\supit{b}INAF-Istituto di Astrofisica Spaziale e Fisica Cosmica,
Sezione di Milano, Via E. Bassini 15,
20133 Milano, Italy; \\
\supit{c}SRON Netherlands Institute for Space Research, Sorbonnelaan 2, 
3584 CA Utrecht, The Netherlands \\
\supit{d}Centre d'Etude Spatiale des Rayonnements, 9, avenue du Colonel 
Roche - BP 4346 31028 Toulouse Cedex 4, Toulouse, France \\
\supit{e}INAF-Istituto di Astrofisica Spaziale e Fisica Cosmica, Sezione di Bologna, via Gobetti 101, 40129 Bologna, Italy \\
\supit{f}Danish Space Research Institute, Juliane Maries Vej 30, DK-2100 
Copenaghen, Denmark \\
\supit{g}Thales Alenia Space, Strada Padana Superiore 290, 20090 Vimodrone, Italy \\ 
}

\authorinfo{Further author information: (Send correspondence to L.Natalucci)\\L.Natalucci: E-mail: Lorenzo.Natalucci@iasf-roma.inaf.it, Telephone: +39 06 4993 4461}

 
  \maketitle

\begin{abstract}
The success of the SWIFT/BAT and INTEGRAL missions has definitely opened 
a new window for follow-up and deep study of the transient gamma-ray sky. 
This now appears as the access key to important progresses in the area of 
cosmological research and deep understanding of the physics of compact 
objects. To detect in near real-time explosive events like Gamma-Ray bursts, 
thermonuclear flashes from Neutron Stars and other types of X-ray outbursts 
we have developed a concept for a wide-field gamma-ray coded mask instrument 
working in the range 8-200 keV, having a sensitivity of 0.4~ph~cm$^{-2}$~s$^{-1}$  
in 1~s (15-150 keV) and arcmin location accuracy over a sky region as wide as 
3~sr. This scientific requirement can be achieved by means of two 
large area, high spatial resolution CZT detection planes made of arrays of 
relatively large ($\sim1$~cm$^2$) crystals, which are in turn read out as matrices 
of smaller pixels. To achieve such a wide Field-Of-View the two units can 
be placed at the sides of a S/C platform serving a payload with a complex 
of powerful X-ray instruments, as designed for the EDGE mission. The two 
units will be equipped with powerful signal read out system and data 
handling electronics, providing accurate on-board reconstruction of the 
source positions for fast, autonomous target acquisition by the X-ray 
telescopes.
\end{abstract}


\section{INTRODUCTION} \label{sect:sections}

The relevance of X- and $\gamma$-ray broadband instrumentation has been largely
established during the last decade,  
for the study of thermal and non-thermal physical processes in
astrophysical sites, as well as their tight interaction spanning a very broad 
region of the electromagnetic spectrum (from the keV to the TeV range). 
Still very difficult 
remain the investigations of transient phenomena mainly due to the need of fast
follow up. The sensitivity of wide field telescopes is intrinsically 
worse than what is needed for dedicated, sophisticated spectroscopy
measurements and the time scale of the events to be studied is often in the
range of seconds or minutes. Especially 
after the results of BeppoSAX (Boella et al, 1987) and SWIFT (Gehrels et al. 2004), 
it is recognized that important 
achievements can be obtained by short-term reaction to transient  
events like Gamma-Ray Bursts (GRBs), outbursts from compact objects and  
thermonuclear flashes from accreting Neutron Stars (NS).

Despite these phenomena have been known for long, 
new types of transient sources
are appearing in most recent years. Among them, the Supergiant Fast X-ray 
Transients are a new class showing hour-long, hard X-ray  
outbursts (Sguera et al. 2006). These are accreting objects associated to High
Mass X-ray Binaries with a Supergiant companion. Their growing number (despite
the difficulty of detecting them) is suggesting they are a dominant class of
binaries in our Galaxy, very interesting as they could be the progenitors of
double NS binaries or NS/Black Hole binaries. Other recent, interesting class
is that of Anomalous X-ray Pulsars (AXP), which have been discovered
to undergo short, extremely hard outbursts (Kaspi 2007, Sguera et al. 2007). 
    
Flaring behaviour of Galactic compact sources at time scales of hours 
to min is, therefore, a rather established but poorly studied phenomenon. 
Other fast and strong events have been found in 
Black Hole (BH) systems, like rapid flux variability of V4641 SGR 
(e.g. Wijnands \& Van der klis 2000) and possibly GRB~070610, likely to be  
associated to a stellar black hole (Kasliwal et al. 2007). We also list  
giant outbursts of Cyg X-1 (Golenetskii et al. 2003) and bright flares from 
XTE J1650-500 (Tomsick et al 2003). These are probably not related to disk instability 
but can be interpreted in many ways, as violent activity of the inner accretion disk 
(e.g. feeding of material into jets), fluctuations of direct wind accretion, presence 
of thick, possibly clumpy absorbers or outflows. High resolution spectroscopy of these 
bright transient events is therefore a tool for important diagnostics, as well as for 
persistent X-ray emission.  In accreting stellar Black Holes, the detection of Fe 
fluorescence lines with “skewed” profile can be explained by  relativistic effects 
at the inner edge of accretion disk. On the other hand, broad and narrow Fe lines 
have been observed from Cyg X-1 while the source was in its “intermediate” state, 
which may be interpreted as originating from reflection on both inner region and disk 
(Miller et al. 2004). 
 
Fine X-ray spectroscopy is also of key importance in the study of NS.  The measurement 
of the gravitational redshift in observations of X-ray bursts, through detection of 
redshifted lines can lead to determination of the NS mass-radius relationship, hence 
constraining the equation of state of  the dense nuclear matter.  An  
important class of thermonuclear bursts are the “superbursts”, i.e. very long ($\sim$~hours) 
events due to unstable burning of carbon and/or other products of the hydrogen/helium 
burning occurring on the NS surface. Despite these events are $\sim$~1000 times more 
energetic than normal type-I bursts, only a few of them are known, while it is very 
important to study their frequency and characteristics as a function of accretion rate 
(see e.g. Sthromayer \& Bildsten 2003). Discrete components by Fe have been found 
in the spectrum of a superburst by 4U1820-30 (Sthromayer \& Brown 2002).
 
We have developed a concept for an instrument exploiting a very wide region of the sky, 
to serve as an  
affordable, sensitive and moderate resource consuming payload complement  
for re-pointing capability in a fast-slewing satellite. 
The former success of the BeppoSAX/WFC (Jager et al. 1997)
in the study of GRB and transient sources has motivated us to develop similar
concept where  
source positioning is obtained by the use of coded masks, proven to be very
successful also in the case of SWIFT and INTEGRAL (Winkler et al.~2003). 
This study, as reported
hereafter has been targeted to the EDGE satellite (den Herder et al. 2007),
to be flown on Low Earth Orbit (LEO). EDGE is 
a proposal for a class-M mission in the framework of the ESA Cosmic Vision 
Programme and carries a payload complex consisting
of powerful X-ray spectroscopy and imaging telescopes, a Wide Field 
Monitor (WFM, described in this work) 
and a GRB detector.
 
EDGE looks as the ideal mission to exploit the above type of transient phenomena. 
Rapid response to 
X-ray transient events and capability of high resolution spectroscopy are the key to 
probe the extreme relativistic regime of the inner regions in BH accretion disks and 
physical conditions at the neutron star surface, e.g. by detection of line features 
in superbursts. In such a way the phenomena currently observed sporadically will 
be upgraded to real classes allowing to more detailed and well constrained models.

For an overview of the EDGE scientific objectives and payload see  
den~Herder et al. (2007).
 
\section{DESIGN OVERVIEW} 

 
The proposed design is based on the positioning of two identical units onto a 
payload module platform. The two cameras together are covering a region of the
sky centered on the satellite pointing axis and are placed at opposite sides of
a narrow field instrument complex. The composition of the FOVs of the two cameras and
the large area of the detectors will ensure proper monitoring of a sky region as
large as $\sim3$~sr. Each WFM camera is complemented by a separate scintillator
instrument for high energy extension (up to $\sim1$~MeV).  

As baseline detector type we have chosen CdZnTe (CZT), mainly
due to its well-proven technology and excellent behaviour in space. Alternative 
solutions have been considered as the use of Silicon Drift devices coupled to CsI
(Marisaldi \& Labanti, 2004) and now regarded as possibly improved configuration 
awaiting further
development in the next years. CZT detector technology is under development in 
Italy with a dedicated research and development program (Quadrini et al. 2007)
as well as in many other european countries.   

In figure 1 the baseline design scheme is described for a single camera (side view). The 
Wide-Field-Monitor (WFM) camera consists of one detection plane and one mask plane 
both inclined by an angle of 28 degrees from the S/C vertical axis (which is coincident 
with the viewing direction of the X-ray telescopes). The overall shape of the camera 
is not regular, for ease of payload accomodation (see Fig.~2).
For the coded mask all the available
space has been exploited, whereas the detector size is basically dictated by 
mass and power budget limitations. 
Detector and mask are separated by a distance of 40.5~cm, allowing for a
source location accuracy of $\sim4$~arcmin to be reached for objects brighter
than $\sim10$~$\sigma$ (see Sect.~3.1 for more details).

   \begin{figure}
   \begin{center}
   \begin{tabular}{c}
   \includegraphics[height=11cm]{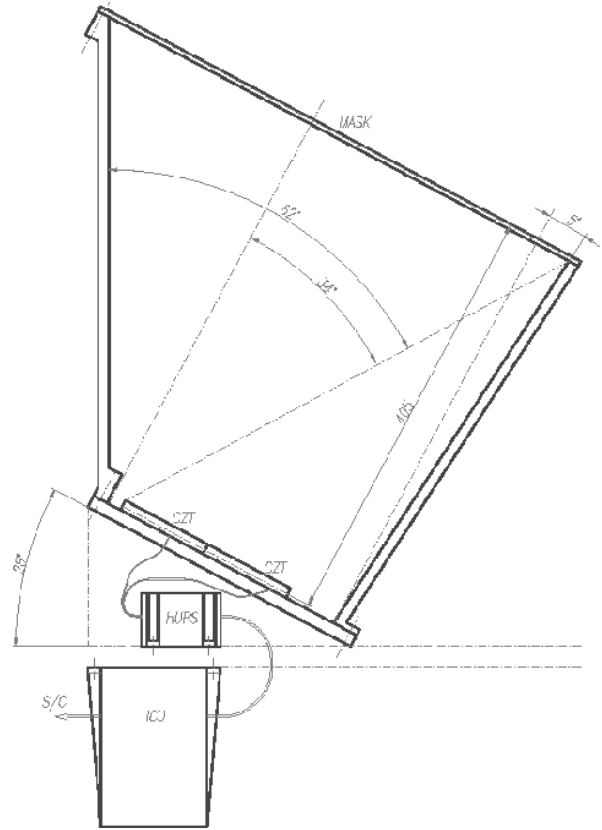}
   \end{tabular}
   \end{center}
   \caption[example]
   { \label{fig:wfm_sideview}
Design overview of one WFM camera (side view). The unit has the detector plane and 
mask plane 
inclined by 28 degrees respect to the S/C pointing axis. The vertical panel on the left 
is on the internal side (close to the X-ray telescope complex) while the oblique 
panel on the right is on the external side of the PLM structure, close to the 
radiator. The Instrument Control Unit is placed below the Payload Module plane, 
and provides the power to 
the detectors via the HVPS box.}
   \end{figure}

   \begin{figure}
   \begin{center}
   \begin{tabular}{c}
   \includegraphics[height=14cm,angle=90]{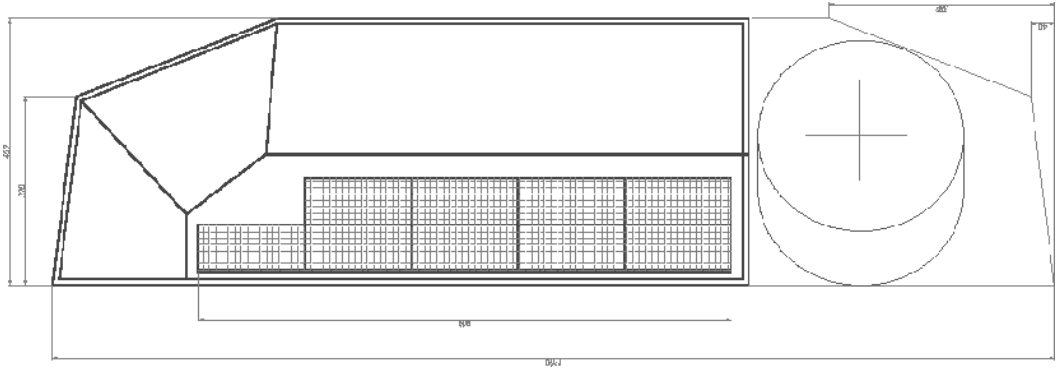}
   \end{tabular}
   \end{center}
   \caption[example]
   { \label{fig:wfm_topview}
Design overview of one WFM camera (top view). A total of 9 CZT modules are assembled 
in two lines: 5 on the internal side and 4 on the external side. The large irregular 
polygonal shape represents the mask pattern placed at 40.5cm from detection plane, 
and connected to the bottom plane by side panels which are acting as a passive
shield. The red cylinder at the right is the GRB detector. All dimensions are 
related to the projection on the S/C PLM plane (note that the camera is inclined 
by 28 degrees).  }
   \end{figure}

The geometric area of the detection plane is approximately 1410~cm$^2$, to ensure 
an active area of 1340~cm$^2$ (the geometric area depends on the final detector 
design, mounting, thermal constraints etc.) and the energy range is such to cover 
effectively a wide band (8-200 keV).  The inclination angle is chosen such as 
to ensure both a large aperture FOV and good matching with the payload allocation 
constraints.  The unit is passively shielded on all sides but the mask aperture, 
and sustained as a whole on the PLM platform by means of dedicated support. The shield 
is obtained by assembly of 5 side panels on which the coded mask support is fixed (by 
means of anchor bolts) and by one bottom shield placed under the CZT plane. No
other collimating devices are present. Since 
the main source of background rate is the diffuse cosmic radiation, there is no  
no real need for active anti-coincidence parts, thus allowing considerable 
weight saving.  
The shield units will be obtained by connecting one or two panels made of W 
and/or graded shield material, 
to be glued onto a CFRP sandwich and reinforced by additional grid structure.

A breakdown of the WFM components is graphically summarized in Fig.~3.

   \begin{figure}
   \begin{center}
   \begin{tabular}{c}
   \includegraphics[height=10cm]{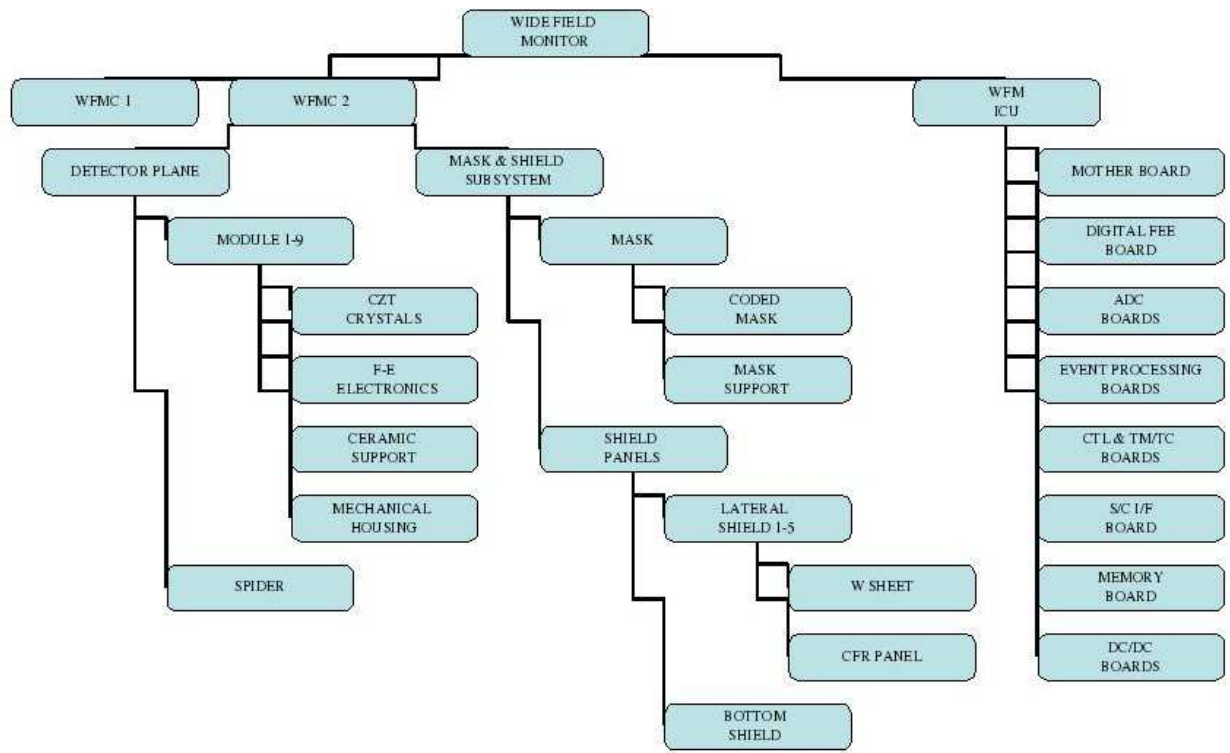}
   \end{tabular}
   \end{center}
   \caption[example]
   { \label{fig:wfm_hwtree}
The WFM hardware tree showing the different subsystems.
}
   \end{figure}
 
\section{CZT ARRAY DETECTOR} \label{sect:sections}

CdZnTe are room temperature, semiconductor devices showing very good performance in terms
of detector efficiency, spectral resolution an broad band coverage. The photon 
interaction cross section is dominated by photoelectric absorption up to 
$\sim200$~keV. CZT crystals can be produced in large number and are also suitable 
to be assembled and integrated to form large area, pixellated detectors. These can
be designed with a large variety of size, thicknesses, and array dimensions.

The electrical design of the detector is crucial to determine its 
performance. Due to the poor mobility of holes in the crystal compared to that
of electrons, the signal amplitude will depend on the depth of 
the energy deposition site. This charge loss can be reconstructed to evaluate
correctly the energy deposition, by recording and analysing the signal shape. 
Moreover the use of segmented electrodes (so-called "small pixel" configuration)  
causes the contribution of holes to the total charge collection to be much reduced,
making the signal amplitude less dependent on the interaction depth. This technique is  
being studied through development of prototypes, including powerful FPGA based digital 
processing unit for
events time sorting,  electronic noise reduction, measurement of signal shape
parameters (see Quadrini et al. 2007). 
 
\subsection{Detector design} 

The detection plane of a single camera is formed by the assembly of 9 identical 
modules, each of 
18~x~9~cm$^2$ (the size is approximated, being slightly dependent on the final number 
of pixels that can be achieved) mounted as shown in Fig.~4.  Each module consists 
of one array of 16~x~8 CdZnTe square crystals with lateral size 10.8mm. As baseline, 
the thickness is at least 2mm (the CZT thickness can be easily selected from 1 
to 10mm as function of the desired energy range and weight budget considerations).  
In turn, each crystal is read out as an array of smaller pixels. First prototypes 
foresee crystals divided in matrix of 4~x~4 and 16~x~16 pixels, each one with its 
own readout chain.  
As baseline, we will consider the basic crystal to be read out as 4~x~4. 
This yields a pixel pitch of 2.7mm and a total of (4x4)x8x16=2048 pixels/module, 
i.e. 18432 pixels for each camera. This choice implies an angular resolution 
of about 35 arcmin 
for a mask placed at a distance of 40.5cm. Potentially a finer sub-division 
(16~x~16 pixels) is also feasible at the expense of additional power and complexity. 
This choice is considered a reasonable compromise between available power budget and 
location accuracy requirement, however the baseline may be hopefully changed if
new technological achievements will make possible to increase the detector 
spatial resolution e.g. by a factor 2 within reasonable power budget. The development
of low power ASICs for space application is promising. Prototypes to be used with 
other semiconductor (GaAs) detectors show power
consumptions as low as 0.8~mW/ch (Bastia et al. 2006). These have 
already been manufactured and tested and we 
are presently considering the possibility of customizing this chip for application 
to CZT detector prototype to be built in the near future.

   \begin{figure}
   \begin{center}
   \begin{tabular}{c}
   \includegraphics[height=4cm]{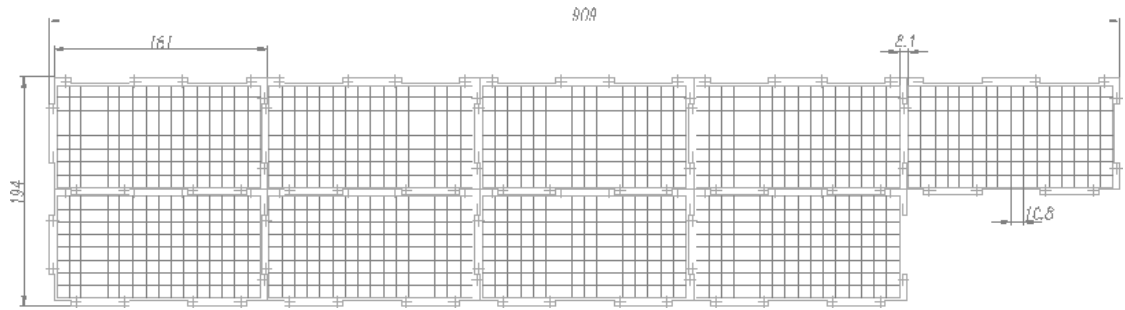}
   \end{tabular}
   \end{center}
   \caption[example]
   { \label{fig:wfm_det}
Schematic of detector assembly for one WFM unit (9 modules, 156~cm$^2$ each). Small red 
squares are the CZT crystals. Each crystal is in turn divided into 4x4 pixels.
}
   \end{figure}

In the current design the FEE is made of 32x32 ASIC readout system, with 2 
ASICs/module mounted below the CZT and within the spider acting as detector support. 
The space between detection plane and S/C PLM plane can be used to host also the 
HVPS units for the detector modules.  Single crystals are mounted within egg-crate 
cells hosting 4x4 crystals each (8 cells/module) as shown in Fig.~1.  Each module 
is in turn controlled by the DFEE for signal acquisition and conditioning, 
performing ADC conversion, background rejection and event reconstruction. Most  
applications will be implemented in the Instrument Control Unit (ICU, see below).

As a baseline, each module will be protected by a thin cover for optical light 
rejection. A thermal blanket 
for the mask can be foreseen in case of threshold energy higher than $\sim7$~keV. 
If the lower level threshold has to be reduced, this blanket may have to be 
replaced by a removable cover.

   \begin{figure}
   \begin{center}
   \begin{tabular}{c}
   \includegraphics[height=13cm]{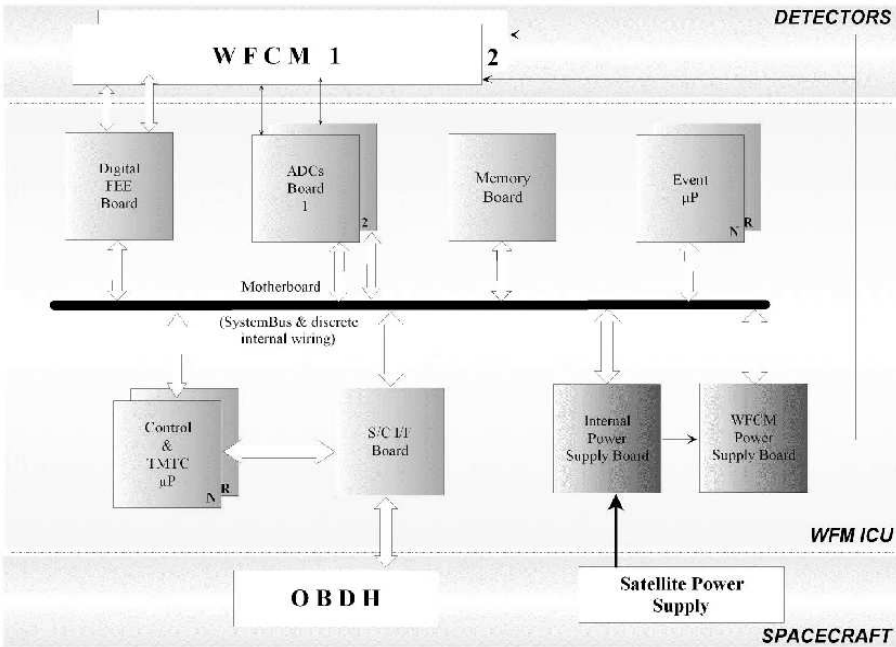}
   \end{tabular}
   \end{center}
\vspace{-5cm}
   \caption[example]
   { \label{fig:icu_architecture}
Instrument Control Unit (ICU) functional block diagram. The S/C interfaces 
for both data/commands and power supply are also indicated. 
}
   \end{figure}

\section{INSTRUMENT CONTROL UNIT}

A single ICU serving both WFM cameras controls the S/C interface, TM/TC and provides 
the processing for trigger, validation and positioning of $\gamma$-ray 
transient events using data from both cameras (in Fig.~5 is shown a functional 
diagram) and from the GRB detector. 
The WFM ICU houses a total of 12 boards, two of which
are redundant microprocessor units.  

The ICU is mounted below the PLM plane as shown schematically in Fig.~1. 
Its approximate size is 26x32x20~cm$^3$.  In Fig. 5 is shown the functional diagram 
with the ICU components and I/Fs, as well as identification of the tasks 
performed. 

Data compression and handling of the different operating modes will be performed
to ensure smooth operations and reliable trigger mechanism. This will be based on 
the elaboration of count rates on different time scales and energy intervals from
different detector sections. A coincidence with the GRB detector 
(possibly via the S/C interface) will be 
normally required, to avoid the largest possible number of false triggers. 
The source positioning will be obtained by flat-fielding, cross-correlation and
peak search methods. Criteria for source validation and selection will be defined by
dedicated future study. This will take advantage in particular, from the expertise
in the development of the GRB alert and fast localization procedures 
on board the AGILE satellite (Tavani et al. 2006). 

\section{RESOURCE BUDGETS}
The mass and power resources needed by the WFM have been evaluated. The weight 
for each WFM camera is approximately 51 kg, shared between detector ($\sim19$~kg), 
mask ($\sim10$~kg) and shield ($\sim22$~kg). Adding 8~kg to be allocated for 
the ICU the total
instrument weight is 111 kg (133 kg including 20\% contingency). 
The total power, as primary line at S/C level is estimated to be approximately
127 W (140 kg with 10\% contingency) by using a basic figure of 1.2 mW/channel.

The long time averaged data production rate is dominated by the count rate intensity 
when observing with the Galactic Plane in the FOV (up to ~5000 c/s per camera). 
Data are stored on board each orbit and then trasmitted. 
In case no onboard data selection or compression scheme are considered,
and assuming that 10\% of the time is available for download (9~min out
of 90~min orbit period)
the total telemetry demand is estimated to be $\sim5$~Mbits/s.   
In case the threshold is lowered to e.g 5~keV or less  
the count rate will increase by about a factor two. The above estimated value
is at least a factor 3 more demanding than the current limit requirement of 
1.5~Mbits/s (based on a total data rate of 5~Mbits/s for the whole EDGE
mission). Therefore it is foreseen that at least part of the data 
will be not trasmitted as photon by photon but histogrammed on board. In this
case, it is required to extend the event processing and background 
rejection capability, as implemented in the ICU. 

\section{PERFORMANCE PREDICTION}

\begin{table*}[]
\caption[]{ Instrument performance parameters vs design requirements. The 
"model" column represents the actual performance of the baseline configuration
of the WFM. 
}
\hspace{2.2cm}
\vspace{-3mm}
\begin{flushleft}
\begin{tabular}{l|c|c|c|l}
\hline

\vspace{-3mm}\\
  Parameter & Requirement & Model & Goal & Comment   \\
\hline
Energy Range (keV) & 8-200 & 8-200 & 5-200 & Region with on axis effective area \\
 & & & & $>20$\% of max. \\ 
Field-Of-View  & 2.5 sr & 2.5 sr & 3 sr & Region with average effective area \\
 & & & & $>300$~cm$^{2}$  \\
Energy Resolution & 5\% & 3\% & 3\% & Values at 100 keV  \\
Effective Area 10-100 keV & 1000~cm$^{2}$  & 1100~cm$^{2}$ & 1200~cm$^{2}$ & 
 Determined on S/C pointing axis \\ 
Angular Resolution & 35~arcmin & 34~arcmin & 18~arcmin & FWHM resolution \\
Location Accuracy & 4~arcmin & 4~arcmin & 2~arcmin & SNR $>10$~$\sigma$  \\
Time Resolution & 10~ms & 10~$\mu$s & 10~$\mu$s & \\
Max. Count Rate & 5000 c/s & 5000 c/s & 5000 c/s & Max. value estimated for \\
 & & & & Galactic Plane in FOV \\
S/W Processing Time & 20~s & $<20$~s & 5~s & Time necessary for burst localization \\  
Continuum Sensitivity & 0.5 & 0.4 & 0.4 & 15-150 keV band, 1s integration time \\
(ph~cm$^{-2}$~s$^{-1}$) & & & & \\
\hline
\end{tabular}
\end{flushleft}
\label{tab:spectral}
\end{table*}

In figure 6 is shown the effective area of the WFM for a source in the direction 
coincident with the satellite pointing. This includes the area projections, the 
total detection efficiency and mask transparency and open fraction (assumed 50\%) 
as well 
as the efficiency loss due to image reconstruction (finite spatial resolution of 
detection plane). 

For the purpose of the WFM however, an important performance issue is how the  
effective area changes with varying offset direction. 
To evaluate the capability of the WFM to 
detect a source in a large enough FOV, we compute the average exposed area 
for sources shining at a given offset angle respect to the S/C pointing direction.
The average is computed along a circle corresponding to the offset.
This is obtained by composition of the two FOVs related to the two 
different cameras.  The plot in Fig.~7 shows that the effective area 
exposed to a source within 2.5~sr is always greater than 300~cm$^2$, which 
is evaluated as the limit to detect bursts with fluences higher than
10$^{-6}$~erg/cm$^2$ (the lower limit 
requirement for bright GRBs). For bursts brighter than two times this limit the 
useful region increases to 3.7~sr.  

The sensitivity of the WFM in the 10-200 keV region is very similar to the
INTEGRAL/IBIS one. 
Below 100 keV it is estimated as 2-3~mCrab in 10$^5$s, which translates 
to $\sim1$~Crab in 1 sec (the sensitivity vs short events is most relevant
for the WFM). In summary the instrument will have a sensitivity similar to IBIS/ISGRI 
with an expected improvement for E$>150$~keV,  
due to the better background conditions in low earth orbit. 

In Table I we report the basic performance parameter as evaluated today, 
compared to the performance requirements of the EDGE mission.

   \begin{figure}
   \begin{center}
   \begin{tabular}{c}
   \includegraphics[height=6cm]{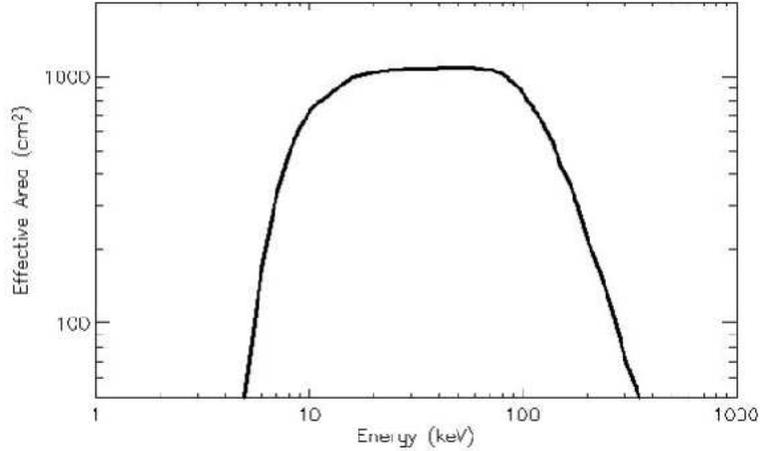}
   \end{tabular}
   \end{center}
   \caption[example]
   { \label{fig:collimator}
On axis effective area of the Wide-Field-Monitor. It includes contribution from
both cameras and refers to a direction coincident with the S/C viewing axis.
}
   \end{figure}

   \begin{figure}
   \begin{center}
   \begin{tabular}{c}
   \includegraphics[height=8cm]{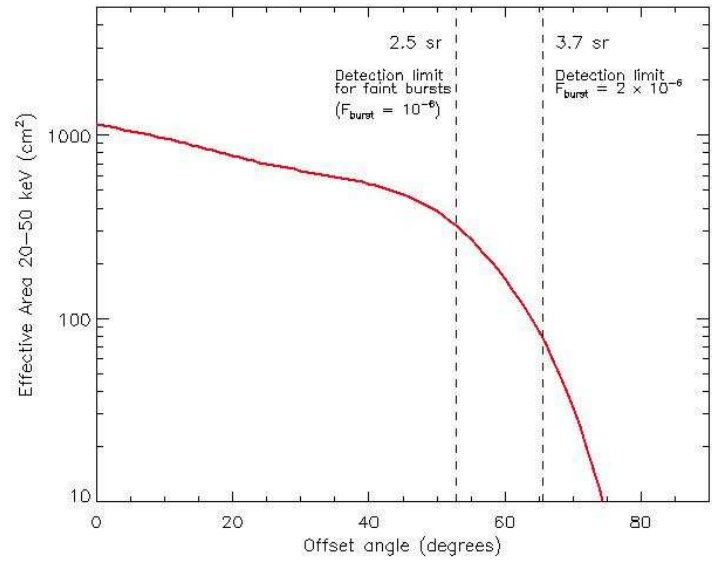}
   \end{tabular}
   \end{center}
   \caption[example]
   { \label{fig:collimator}
The effective area of the two WFM units, shown versus
the offset angle to the S/C pointing axis. Indicated by the dashed
lines are the offset values, corresponding to the minimum effective area needed 
to detect bursts with fluences of 10$^{-6}$~erg/cm$^2$ and 
$2\times10^{-6}$~erg/cm$^2$. 
The fluence limit for faint GRBs corresponds to the design requirement of the
EDGE mission for the detection of GRBs for the WHIM studies.
}
   \end{figure}




\begin{thebibliography}{1}  

\bibitem{Bas06}
Bastia, P., Bertuccio, G., Borghetti, F. et al.,
"A complete Read-Out ASIC for Use with large Pixel X-ray Detector Array",
ESA-SP 630, p.28, 2006

\bibitem{Boe97}
Boella, G., Butler, R. C., Perola, G. C. et al.,
"BeppoSAX, the wide band mission for X-ray astronomy",
A\&A Suppl., 122, 299, 1997

\bibitem{Den07}
den~Herder, J.W., Piro, L., Ohashi, T. et al., 
"EDGE: Explorer of Diffuse Emission and Gamma-ray Burst Explosions",
these Proceedings, Conf. 6688, 2007

\bibitem{Geh04}
Gehrels, N., Chincarini, G., Giommi, P. et al., 
"The SWIFT Gamma-Ray Burst Mission",
ApJ 611, 1005, 2004 

\bibitem{Gol03}
Golenetskii, S., Aptekar, R., Frederiks, D., Mazets, E., Palshin, V., 
"Observations of Giant Outbursts from Cygnus X-1",
ApJ 596, 1113, 2003

\bibitem{Jag97}
Jager, R., Mels, W. A., Brinkman, A. C., et al., 
"The Wide Field Cameras onboard the BeppoSAX X-ray Astronomy Satellite",
A\&A 125, 557, 1997

\bibitem{Kas07}
Kasliwal, M.M., Cenko, S.B., Kulkarni, S.R. et al.,
"GRB070610: A Curious Galactic Transient"
ApJ submitted, arXiv:0708.0226v1 [astro-ph], 2007 

\bibitem{Ksp07}
Kaspi, V.M.,  
"Recent Progress on Anomalous X-ray Pulsars", 
Astrophysics \& Space Science, Vol.~308, No. 1-4, p.1, 2007

\bibitem{Mar04}
Marisaldi, M., Labanti, C., Bulgarelli, A. et al., 
"A Position Sensitive Gamma-Ray Detector Based on Silicon Drift 
Detectors Coupled to Scintillators for Application in the MEGA Compton 
Telescope", Proc. IEEE NSS Conf., 2004

\bibitem{Mil04}
Miller, J.M., Raymond, J., Fabian, A.C. et al., 
"Chandra/High Energy Transmission Grating Spectrometer Spectroscopy 
of the Galactic Black Hole GX 339-4: A Relativistic Iron Emission Line 
and Evidence for a Seyfert-like Warm Absorber",
ApJ 601, 450, 2004

\bibitem{Qua07}
Quadrini, E.M, Uslenghi, M., Alderighi, M. et al., 
"Spectroscopic CZT detectors development for X and Gamma-Ray 
Imaging Instruments", 
these Proceedings, Conf. 6686, 2007 

\bibitem{Sgu06}
Sguera, V., Bazzano, A., Bird, A.J. et al., 
"Unveiling Supergiant Fast X-Ray Transient Sources with INTEGRAL",
ApJ 646, 452, 2006

\bibitem{Sgu07}
Sguera, V., Bazzano, A., Bird, A.J. et al.,
"INTEGRAL high energy detection of the transient IGR J11321-5311",
A\&A, in press, arXiv:0704.2737v1 [astro-ph], 2007

\bibitem{Str03}
Strohmayer, T.E  \& Bildsten, L., 
"New views of thermonuclear bursts", in "Compact Stellar X-ray Sources",
eds. W.H.G. Lewin and M. van der Klis, Cambridge, Cambridge University Press, 2003

\bibitem{Str02}
Strohmayer, T.E., \& Brown, E.F., 
"A Remarkable 3 Hour Thermonuclear Burst from 4U 1820-30",
ApJ 566, 1045, 2002

\bibitem{Tav06}
Tavani, M., Barbiellini, G., Argan, A., 
"The AGILE Mission and its Scientific Instruments",
Proc. SPIE 2006, vol. 6266, 2006

\bibitem{Tom03}
Tomsick, J.A., Kalemci, E., Corbel, S., Kaaret, P., 
"X-Ray Flares and Oscillations from the Black 
Hole Candidate X-Ray Transient XTE J1650-500 at Low Luminosity", 
ApJ 592, 1100, 2003

\bibitem{Wij00}
Wijnands, R.  \& Van der Klis, M., 
"The Rapid X-Ray Variability of V4641 Sagittarii 
(SAX J1819.3-2525 = XTE J1819-254)",
ApJ 528, L93, 2000

\bibitem{Win03}
Winkler, C., Courvoisier, T.~J-L., Di Cocco, G. et al., 
"The INTEGRAL Mission"
A\&A 411, L1, 2003


\end{thebibliography}
\end{document}